\documentclass[twocolumn,showpacs,preprintnumbers,amsmath,amssymb]{revtex4}

\usepackage{graphicx}
\usepackage{dcolumn}
\usepackage{bm}

\begin{document}

 \title{Three body resonances in two meson-one baryon systems}

\author{A. Mart\' inez Torres}
\email{amartine@ific.uv.es}
\author{K. P. Khemchandani}%
\author{E. Oset}%
\affiliation{ Departamento de F\' isica Te\' orica and IFIC, Centro Mixto Universidad de Valencia-CSIC, Institutos de Investigaci\'on
de Paterna, Aptd. 22085, 46071 Valencia, Spain.
}


\date{\today}

\begin{abstract}

We report four $\Sigma$'s and three $\Lambda$'s, in the 1500 - 1800 MeV region, as two meson - one baryon S-wave $(1/2)^+$ resonances.  We solve Faddeev equations in the coupled channel approach. The invariant mass of one of the meson-baryon pairs and that of the three particles have been varied and peaks in the squared three body $T$-matrix have been found very close to the existing $S$ = -1, $J^P= 1/2^+$ low lying baryon resonances. The input two-body
$t$-matrices for meson-meson and meson-baryon interaction have been calculated by solving the Bethe-Salpeter equation with the potentials obtained in the chiral unitary  approach.

\end{abstract}

\maketitle

Hyperon physics is still relatively unexplored and there are many open problems to be studied. The hyperon resonances, for example, have a poor status as compared to the nucleon ones \cite{pdg}. Though some of
them, like the low lying $J^P= 1/2^-$ resonances, $\Lambda(1405)$, $\Lambda(1670)$ $\dots$, can be represented as dynamically generated and relatively well understood states within the unitary chiral models, the low lying $J^P= 1/2^+$ domain remains far less understood, both experimentally and theoretically. For instance, quark models seem to face difficulties in reproducing properties of the resonances in this sector \cite{riska}. The neat reproduction of the low lying $1/2^-$ states in the $S$-wave meson-baryon interaction, using chiral dynamics, suggests that the addition of a pseudoscalar meson in S-wave could lead to an important component of the structure of the $1/2^+$ resonances. There exist results which hint towards this possibility, e.g., in \cite{longcola} it was found that the two meson cloud gave a sizable contribution to the mass in the spectrum of the $1/2^+$ baryon antidecuplet. Chiral dynamics has been used earlier in the context of the three nucleon problems, e.g., in \cite{epelbaum}. In this article we present the first study of two meson - one baryon systems where chiral dynamics is applied to solve the Faddeev equations. As shall be described below, our calculations for the total $S$ = -1 reveal peaks in the  $\pi \bar{K} N$ system and its coupled channels which we identify with the resonances $\Sigma(1770)$, $\Sigma(1660)$, $\Sigma(1620)$, $\Sigma(1560)$, $\Lambda(1810)$ and $\Lambda(1600)$.

We start by taking all combinations of a pseudoscalar meson of the $0^-$ SU(3) octet and a baryon of the $1/2^+$ octet which couple to $S=-1$ with any charge. For some quantum numbers, the interaction of this two body system is strongly attractive and responsible for the generation of the two $\Lambda(1405)$ states \cite{jido} and other $S$ = -1 resonances. We shall assume that this two body system formed by $\bar{K}N$ and coupled channels remains highly correlated when a third particle is added, in the present case a pion. Altogether, we get twenty-two coupled channels for the net charge zero configuration: $\pi^0 K^- p$, $\pi^0\bar{K}^0 n$, $\pi^0\pi^0\Sigma^0$, $\pi^0\pi^+\Sigma^-$, $\pi^0\pi^-\Sigma^+$, $\pi^0\pi^0\Lambda$, $\pi^0\eta\Sigma^0$, $\pi^0\eta\Lambda$, $\pi^0 K^+\Xi^-$, $\pi^0 K^0\Xi^0$, $\pi^+ K^- n$, $\pi^+\pi^0\Sigma^-$, $\pi^+\pi^-\Sigma^0$, $\pi^+\pi^-\Lambda$, $\pi^+\eta\Sigma^-$, $\pi^+ K^0\Xi^-$, $\pi^-\bar{K}^0 p$, $\pi^-\pi^0\Sigma^+$, $\pi^-\pi^+\Sigma^0$, $\pi^-\pi^+\Lambda$, 
$\pi^-\eta\Sigma^+$, $\pi^- K^+ \Xi^0$. We assume the correlated pair to have a certain invariant mass, $\sqrt{s_{23}}$, and the three body $T$-matrix is evaluated as a function of  this mass and the total energy of the three body system.

The input required to solve the Faddeev equations, i.e., the two body $t$-matrices for the meson-meson and meson-baryon interactions have been calculated by taking the lowest order chiral Lagrangian following
\cite{npa,angels,bennhold,Inoue} and using the dimensional regularization of the loops as done in
\cite{ollerulf,bennhold}, where a good reproduction of scattering amplitudes and resonance properties was found. Instead, a cut off could also be used to regularize the loops as shown in \cite{angels,ollerulf}. Improvements introducing higher order Lagrangians have been done recently, including a theoretical error analysis \cite{boraulf} which allows one to see that the results with the lowest order Lagrangian fit perfectly within the theoretical allowed bands.

A shared feature of the recent unitary chiral dynamical calculations is the on-shell factorization of the potential and the $t$-matrix in the Bethe-Salpeter equation \cite{npa,angels,ollerulf,Nieves:1999bx,carmen,hyodo,borasoy}, which is
justified by the use of the N/D method and dispersion relations \cite{nsd,ollerulf}. Alternatively, one can see that the off-shell contributions can be reabsorbed into renormalization of the lower order terms \cite{npa,angels}. We develop here a similar approach for the Faddeev equations.

The full three-body $T$-matrix can be written as a sum of the auxiliary $T$-matrices $T^1$, $T^2$ and $T^3$ \cite{Faddeev}
\begin{equation}
T=T^1+T^2+T^3
\end{equation}
where $T^i$, $i=1$, $2$, $3$, are the normal Faddeev partitions, which include all the possible interactions contributing to the three-body $T$-matrix with the particle $i$ being a spectator in the last interaction.
The Faddeev partitions satisfy the equations
\begin{equation}\label{eq:Tiorig}
T^i=t^i\delta^3(\vec{k}^{\,\prime}_i-\vec{k}_i)+ t^i g^{ij}T^j + t^i g^{ik}T^k ,
\end{equation}
where $\vec{k}_i$ ($\vec{k}^{\,\prime}_i$) is the initial (final) momentum of the ith particle in the global center of mass system, $t^i$ is the two-body $t$-matrix for the interaction of the 
pair $(jk)$ and $g^{ij}$ is the three-body propagator or Green's function, with $j \neq k \neq i$ = 1, 2, 3 

Iterating Eq.(\ref{eq:Tiorig}) and removing the term with $\delta^3(\vec{k}^{\,\prime}_i-\vec{k}_i)$, which correspond to a ``disconnected diagram''\cite{Noyes}, will give
\begin{eqnarray}\label{expand}
T^i &=& t^i g^{ij} t^j + t^i g^{ik} t^k + t^i g^{ij} t^j g^{jk} t^k + t^i g^{ij} t^j g^{ji} t^i+\nonumber\\  
&&+t^i g^{ik} t^k g^{kj} t^j + t^i g^{ik} t^k g^{ki} t^i + \cdots 
\end{eqnarray}

\begin{figure}[ht]
\includegraphics[width=0.5\textwidth] {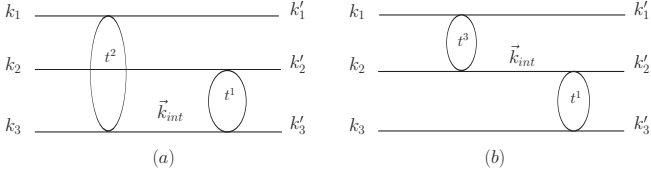}
\caption{\label{fig1} Diagrammatic representation of the terms (a) $t^1 g^{12} t^2$
(b) $t^1 g^{(13)} t^3$.}
\end{figure}

\begin{figure}[ht]
\includegraphics[scale=0.6] {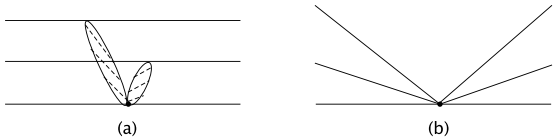}
\caption{\label{fig2} The origin of the three body forces (a) due to cancellation of the propagator in Fig.\ref{fig1}(a) with the off-shell part of the chiral amplitude, (b) at the tree level.}
\end{figure}

The first two terms of the Eq.(\ref{expand}), for the case $i$=1, are represented diagrammatically in Fig.\ref{fig1},
where the $t$-matrices are required to be off-shell. However, the chiral amplitudes, which we use,
can be split into an ``on-shell'' part (obtained when the only propagating particle of the diagrams, labeled with $\vec{k}_{int}$ in Fig.\ref{fig1}, is placed on-shell), which depends only on the c.m energy of the interacting pair, and an off-shell part proportional to the inverse of the propagator of the off-shell particle. This term would cancel the particle propagator (for example that of the 3rd particle in the Fig.\ref{fig1}a) resulting into a three body force (Fig.\ref{fig2}a). In addition to this,
three body forces also stem directly from the chiral Lagrangians \cite{Felipe} (Fig.\ref{fig2}b).

We find that the sum of the off-shell parts of all the six $t^i g^{ij} t^j$ terms, together with the contribution from Fig.\ref{fig2}(b) cancels exactly if the SU(3) limit is considered and the momentum transfer for the baryon is assumed to be small. In a realistic case we find this sum to be smaller than 5$\%$ of the total on-shell contribution. Hence, only the on-shell part of the two body (chiral) $t$-matrices is significant. The diagrams in Fig.1 can then be expressed mathematically (reading the diagrams from right to left as a convention) as $t^1 g^{12} t^2$ and $t^1 g^{13} t^3$, 
respectively, where the $t^i$-matrices depend only on the center of mass energy of the interacting particles. 

The $t^ig^{ij}t^j$ terms correspond to the situation where there are no loops 
and hence the $g^{ij}$ propagators are written in terms of the on-shell variables
\begin{equation}
g^{ij}=\Bigg(\prod_{r=1}^D\dfrac{N_r}{2E_r}\Bigg)\dfrac{1}{\sqrt{s}-E_i
(\vec{k}^\prime_i)-E_j(\vec{k}_j)-E_k(\vec{k}^\prime_i+\vec{k}_j)+i\epsilon}\nonumber
\end{equation}
with $\sqrt{s}$ being the total energy in the global CM system. $E_l=\sqrt{\vec{k}^2_l+m^2_l}$ is the total energy of the particle $l$ and $N_l$ is a normalization constant ($N_l=1$ for the meson-meson interaction and $N_l=2M_l$ for the meson-baryon interaction, where $M_l$ is the corresponding baryon mass) and $D$ is the number of particles propagating between two consecutive interactions.

These propagators (and all other angle dependent expressions in the formalism) are projected in S-wave. A proper Lorentz boost has been made for transformation of the momenta from the center of mass frame of two particles to the global center of mass frame whenever needed. A technical remark is here in order: to avoid the evident divergence in these on-shell propagators at the threshold of a channel and to continue the Faddeev equations analytically below the threshold, we fix the momentum of the particle to a minimum value, $P_{min} =$ (for example, 50 MeV). It should be mentioned that the results are almost insensitive to a change in the value of the $P_{min}$, since a change in $P_{min}$ of $\sim$ 40- 50$\%$ results into
a shift in the peak position by less than 2 MeV.

\begin{figure}[ht]
\includegraphics[scale=0.40] {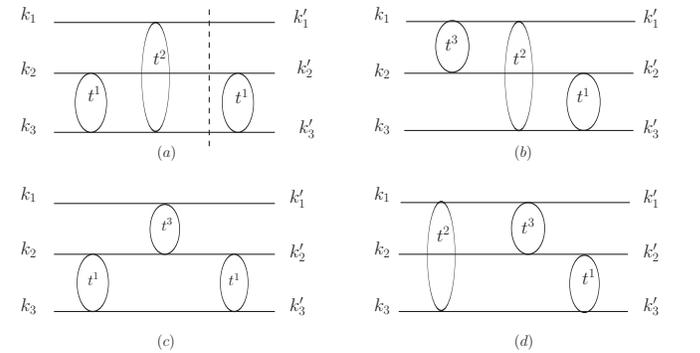}
\caption{\label{fig3} Different diagrams involving three pair interactions corresponding  to the $T^1$ partition.}
\end{figure}

The first term with a non trivial structure, from the point of view of the on-shell factorization of the $t$-matrices in the Faddeev equations, is the one involving three successive pair interactions, where a loop function of three particle propagators appears for the first time. We show the diagrams with such a structure for the $T^1$ partition in Fig.\ref{fig3}(a-d).

We write all these $t^i g^{ij} t^j g^{jk} t^k$ terms as $t^i G^{ijk}(t^j g^{jk} t^k)|_{on-shell}$  where
\begin{equation}
G^{i\,j\,k}=\int\dfrac{d^3 k^{\prime\prime}}{(2\pi)^3}\dfrac{N_l}
{2E_l}\dfrac{N_m}{2E_m}\dfrac{F^{i\,j \,k}(\sqrt{s},\vec{k}^{\prime\prime}
)}{\sqrt{s_{lm}}-E_l(\vec{k}^{\prime\prime})-E_m(\vec{k}^{\prime\prime})
+i\epsilon}\label{eq:G}
\end{equation}
with $i \neq j$, $j \neq k$, $i\neq l \neq m$ and $\sqrt{s_{lm}}$ is the invariant mass of the $(lm)$ pair.
Eq.(\ref{eq:G}) consists of the two particle propagator in the first cut (shown as a dashed line in Fig.3(a) as an example) and the $F^{i\,j\,k}$ function, which is defined as
\begin{equation}\label{offac}
F^{i\,j\,k} = t^{j}(\sqrt{s_{int}} (\vec{k}^{\prime\prime})) \Biggr( \dfrac{g^{jk}|_{off-shell}}{ g^{jk}|_{on-shell}} \Biggr) [ t^{j}(\sqrt{s_{int}} (\vec{k}_{j^\prime})) ]^{-1}.
\end{equation}
$s_{int} (\vec{k}^{\prime\prime}) = s - m_j^2 - 2 \sqrt{s} E_j (\vec{k^{\prime\prime}})$ denotes the invariant mass required to calculate the  $t^j$-matrix in the loop integral of $G^{ijk}$. The term $g^{jk}|_{on-shell}^{-1}[ t^{j}(\sqrt{s_{int}} (\vec{k}_{j^\prime})) ]^{-1}$ appearing in $F^{i\,j\,k}$ of Eq.(\ref{offac}) cancels the $(t^j g^{jk})|_{on-shell}$ of $t^i G^{ijk} (t^j g^{jk} t^k)|_{on-shell} $ and produces the $t^i g^{ij} t^j g^{jk} t^k$ term with $g^{ij} t^j$ depending on the proper off-shell variable, $k^{\prime\prime}$ of the loop. This procedure allows us to render the Faddeev equations into a set of algebraic equations, as we see below. To regularize the integrals of Eq.(\ref{eq:G}) we shall use the cut off as in \cite{npa,angels} which is of the order of 1 GeV. The results are rather insensitive to this cut off, due to the convergence of these loops which involve three propagators.

So far we have discussed diagrams with two or three $t$-matrices. It has been shown that the introduction of a third interaction replaces the propagator $g$ by a loop function $G$. The formalism is further developed by  making the same substitution whenever a new interaction is added.

The Faddeev partitions of Eq.(\ref{expand}) in this prescription can be re-written as 
\begin{eqnarray}
T^i&=&\lbrace t^i g^{ij} t^j + t^i G^{ijk} t^j g^{jk} t^k + t^i G^{iji} t^j g^{ji} t^i + \cdots \rbrace\nonumber \\
&+& \lbrace t^i g^{ik} t^k + t^i G^{ikj} t^k g^{kj} t^j + t^i G^{iki} t^k g^{ki} t^i + \cdots \rbrace\nonumber\\
&=& T_R^{ij} + T_R^{ik}
\end{eqnarray}
where the $T^i$ partition has been rewritten in terms of the two infinite series $T_R^{ij}$ and $T_R^{ik}$, which sum all the diagrams with the last two interactions written in terms of $t^i$, $t^j$ and $t^i$, $t^k$, respectively. Hence, we obtain six partitions (which is double of those in the original Faddeev equations);
\begin{eqnarray}\label{trest}
T^{ij}_R&=&t^ig^{ij}t^j+t^i\Big[G^{iji}T^{ji}_R+G^{ijk}T^{jk}_R\Big]
\end{eqnarray}
\noindent
with $i \neq j \neq k$.

It remains to define the invariant masses on which the two body $t$-matrices and the propagators depend.
The expression for the $s_{12}$ and $s_{13}$ obtained from the energy conservation in terms of the external (on-shell) variables is
\begin{equation}
s_{ij} = s + m_k^2 -\dfrac{\sqrt{s} (\sqrt{s} - E_1) (s_{23} + m_k^2 - m_j^2)}{s_{23}}
\end{equation}
with $E_1 = (s - s_{23} + m_1^2 )/(2 \sqrt{s})$.
However, it should be noted that the $t$-matrices $t^{j}(\sqrt{s_{int}}(\vec{k}^{\prime\prime}))$ in the loop (Eq.(\ref{eq:G})) are calculated in terms of a running variable as required.

There are two independent variables in the formalism $\sqrt{s}$, $\sqrt{s_{23}}$, as a function of which we plot the squared $T^*_R$-matrix ($T^*_R = \sum\limits_{ij} ( T_R^{ij} - t^i g^{ij}t^j$ ) ), since the $t^i g^{ij}t^j$ terms
evidently do not have a resonance structure and just provide a background to the amplitudes.

\begin{figure}[ht]
\includegraphics[width=0.6\textwidth]{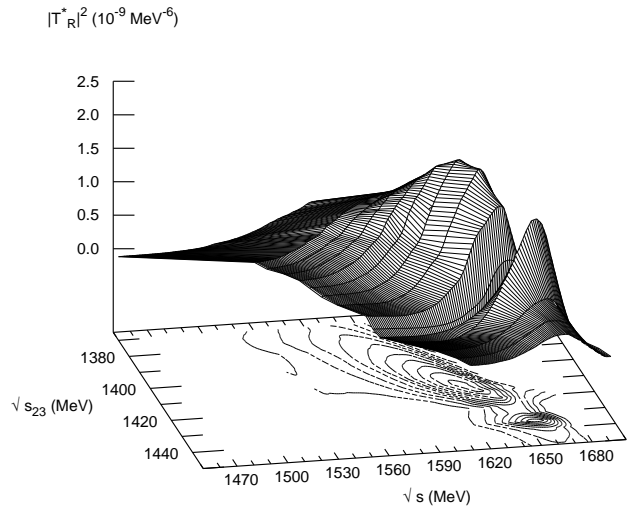}
\caption{\label{fig4} Two $\Sigma$ resonances in the  $\pi \pi \Sigma$ amplitude in $I$ = 1, $I_\pi$ = 2 configuration.}
\end{figure}

We now report the four isospin one states found in our study. In Fig.\ref{fig4}, we show a plot of the squared $T^*_R$-matrix and its projection, for $\pi \pi \Sigma \,\rightarrow \, \pi \pi \Sigma$ in the total isospin $I$ = 1 configuration obtained by keeping the two pions in isospin $I_\pi$ = 2. We see two peaks; one at $\sqrt{s}$ = 1656 MeV with the full width at half maximum $\sim$ 30 MeV and another at $\sqrt{s}$ = 1630 MeV with $\Gamma$ = 39 MeV. We identify the peak at $\sqrt{s}$ = 1656 MeV with the well established $\Sigma(1660 - i100/2)$  \cite{pdg} as a resonance in the $\pi \pi \Sigma$ system, which is a new finding. It is interesting to recall that the excitation of this resonance is claimed in the study of the $K^- \, p \, \rightarrow \, \pi^0\,\pi^0\,\Sigma^0$ reaction \cite{prakhov}. Note that since we are plotting the squared amplitude, which would be proportional to a cross section of a certain process, we can associate our results to the ordinary masses of the $PDG$ and not to the ``pole positions'' also quoted there.

The peak in squared $T^*_R$-matrix observed at 1630 MeV with a width of 39 MeV needs a special attention. The two-star resonance $\Sigma(1620)$ \cite{pdg}, though listed as a $1/2^-$ state, seems to be a very unclear case. The partial wave analysis and the production experiments have been kept separately in \cite{pdg} since it is difficult to know the quantum numbers from the production experiments and if more than one resonance contributes to a single bump. Interestingly, there is a $1/2 ^+$ state found by the partial wave analysis work of Martin et. al. \cite{Martin} in this region. The authors of \cite{Martin} use  a multichannel partial wave analysis of the $\bar{K} N$ data and find a resonance at 1597 MeV. This result has however been listed under the $\Sigma(1660)$ in \cite{pdg}. Another partial wave analysis of the $\bar{K} N \rightarrow \, \Lambda \pi$ reaction made by Armenteros et. al. \cite{Armenteros} find a $1/2^+$ $P_{11}$ resonance at 1610 MeV with a width of 60 MeV. These findings would provide some phenomenological support to our claim of a $1/2 ^+$ $\Sigma$ resonance around $\sim$ 1620 MeV. 

We find two more peaks in the $I=1$ sector; one at $\sqrt{s}$ = 1590 MeV with a width $\sim$ 70 MeV in $I$ = 1, $I_\pi$ = 0 state and another at $\sqrt{s}$ = 1790 MeV with $\Gamma$ = 24 MeV in $I$ = 1, $I_\pi$ = 2 case. The former one supports the existence of the $\Sigma(1560)$ ``bump'', whose spin-parity is unknown \cite{pdg}. Our results would  associate a $1/2 ^+$ to the spin-parity of this resonance. The latter finding supports the one-star $\Sigma(1770)$.

\begin{figure}[t]
\includegraphics[width=0.5\textwidth]{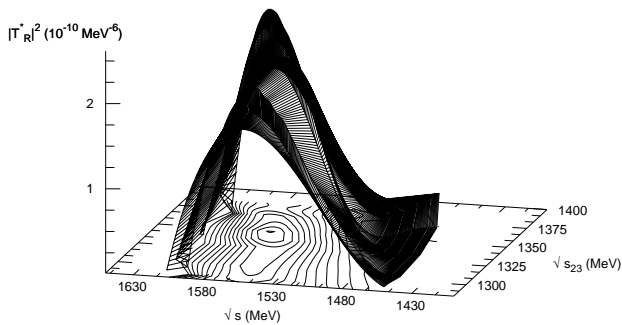}
\caption{\label{fig5}A $\Lambda$ resonance in the $\pi \bar{K} N$ amplitude at 1568 MeV in $I$ = 0, $I_{\pi \bar{K}}$ = 1/2.}
\end{figure}

Next, we discuss the three isospin zero states obtained in these calculations. First we look at states observed in $\pi \bar{K}N$ with $I_{\pi\bar{K}}=1/2$. Two peaks in the $\Lambda(1600)$ MeV region have been found at $\sqrt{s}$ = 1568 with a width of 60 MeV (which is shown in Fig.\ref{fig5}) and at 1700 MeV  with $\Gamma$=136 MeV. One should note that the $PDG$ quotes a mass for the $\Lambda(1600)$ between 1560 MeV and 1700 MeV and the width between 50 MeV and 250 MeV. We should also note the quoting of the $PDG$ concerning this resonance, ``There are quite possibly two $P_{01}$ states in this region''. Our results reinforce this hypothesis.

Finally, in the $\pi \pi \Lambda$ amplitude for the $I$ = 0, $I_\pi$ = 0 configuration we find a similar structure at 1740 MeV with the full width at half maximum being 20 MeV. We identify this peak as the $\Lambda(1810 - i150/2)$ resonance, which is listed as a three-star $1/2^+$ resonance by the particle data group \cite{pdg}. We would like to draw the attention of the reader to the large variation in the peak positions as well as the widths reported by different partial wave analyzes \cite{pdg} for the $\Lambda(1810)$ resonance (the peak position changes from 1750 MeV to 1850 MeV and the width from 50-250 MeV).

We do not find any states with exotic isospin, $I$ = 2,3.

We have also investigated the theoretical uncertainties of the model. We have already mentioned that $P_{min}$ and the cut off in the three particle loop do not play any relevant role. In addition, we have checked the sensitivity of our results to the change in the two body input parameters. We have varied the pion decay constant and the two body cut offs, by about $5\%$, which still guarantees a fair agreement of our two body cross sections with the experimental ones. We find changes in the peak positions by less than $5$ MeV from each source, or $7$ MeV when summed in quadrature. This gives us an idea of the accuracy of our results.

The states obtained are not exotic and their quantum numbers can be reached with just three quarks. What our findings are telling is that in Nature these three quark states unavoidably couple to two mesons and one baryon, and, for the states that we have found, the two meson one baryon  components overcome the weight of the original three quarks seed. This particular nature could be tested experimentally by means of different reactions, among which, the strong three body decay channels and the radiative decays \cite{Michael} should play an important role and deserve further theoretical and experimental studies.

We conclude the discussion by emphasizing that all the low lying ${1/2}^+$ $\Sigma$ and $\Lambda$ resonances in the $PDG$ \cite{pdg}, up to the 1800 MeV energy region, get dynamically generated as two meson-one baryon states
in these calculations. In addition, we predict the quantum numbers of the $\Sigma(1560)$ and also find evidence 
for a  ${1/2}^+$ $\Sigma$ resonance at $\sim$ 1620 MeV. It is rewarding to see that the widths obtained in this work, which correspond to decay into three body systems, are smaller than the total ones to which the two body decay widths also contribute. There would be no contradiction with these two body channels having a smaller weight in the resonance wavefunctions, as implicitly assumed in our study, and having a fair contribution to the total width, since some of the three body channels to which the resonances couple are kinematically closed for decay, and others which are open have a far smaller phase space than that available for two body decay channels.

We would like to thank Profs. C. Hanhart, M. J. Vicente Vacas, V. B. Belyaev and E. O. Alt 
for useful discussions. A. M. T. wishes to acknowledge the support of the Ministerio de Educaci\'on y Ciencia in the program of FPU. This work is partly supported by DGICYT contract
number FIS2006-03438 and the Generalitat Valenciana. This research is  part of
the EU Integrated Infrastructure Initiative  Hadron Physics Project under
contract number RII3-CT-2004-506078. 

\end{document}